\documentclass[%
 reprint,
superscriptaddress,
 amsmath,amssymb,
 aps,
]{revtex4-2}

\usepackage[english]{babel}

\usepackage{amsmath}
\usepackage{graphicx}
\usepackage[colorlinks=true, allcolors=blue]{hyperref}
\usepackage{multirow}
\usepackage{mathrsfs}
\usepackage{upgreek}
\usepackage{lineno}
\usepackage{float}

\begin{document}

\title{Resonant phonons: Localization in a structurally ordered crystal}

\author{Albert Beardo} 
\affiliation{Department of Physics, University of Colorado, Boulder, Colorado 80302}
\affiliation{JILA and STROBE NSF Science and Technology Center, University of Colorado and NIST, Boulder, Colorado 80309}
\author{Paul Desmarchelier}
\affiliation{University of Lyon, CNRS, INSA Lyon, CETHIL, UMR5008. Villeurbane, 69621, France}
\affiliation{Department of Materials Science and Engineering, Johns Hopkins University, Baltimore, Maryland 21218, USA}
\author{Chia-Nien Tsai}
\affiliation{Smead Department of Aerospace Engineering Sciences, University of Colorado, Boulder, Colorado 80303}
\author{Prajit Rawte}
\affiliation{Smead Department of Aerospace Engineering Sciences, University of Colorado, Boulder, Colorado 80303}
\author{\\Konstantinos Termentzidis}
\affiliation{University of Lyon, CNRS, INSA Lyon, CETHIL, UMR5008. Villeurbane, 69621, France}
\author{Mahmoud I. Hussein}
\affiliation{Department of Physics, University of Colorado, Boulder, Colorado 80302}
\affiliation{Smead Department of Aerospace Engineering Sciences, University of Colorado, Boulder, Colorado 80303}

\begin{abstract}
   Phonon localization is a phenomenon that influences numerous material properties in condensed matter physics. Anderson localization brings rise to localized atomic-scale phonon interferences in disordered lattices with an influence limited to high-frequency phonons having wavelengths comparable to the size of a randomly perturbed unit cell. Here we theoretically reveal a new form of phonon localization induced by augmenting a crystalline material with intrinsic phonon nanoresonators with feature sizes that can be smaller or larger than the phonon wavelengths but must be relatively small compared to the phonon mean free paths. This mechanism is deterministic and takes place within numerous discrete narrow-frequency bands spread throughout the full spectrum with central frequencies controlled by design. For demonstration, we run molecular dynamics simulations of all-silicon nanopillared membranes at room temperature, and apply to the underlying thermalized environment narrowband wave packets as an excitation at precisely the frequencies where resonant hybridizations are evident in the anharmonic phonon band structure. Upon comparison to other frequency ranges where the nanostructure does not exhibit local resonances, significant intrinsic spatial phonon localization along the direction of transport is explicitly observed. Furthermore, the energy exchange with external sources is minimized at the resonant frequencies. We conclude by making a direct comparison with Anderson localization highlighting the superiority of the resonant phonons across both sides of the interference frequency limit. 
\end{abstract}


\maketitle

\section{Introduction}

Originally predicted by P. W. Anderson, Anderson localization prevents electron and spin diffusion in random lattices due to wave-like interference effects over distinct scattering paths~\cite{Anderson1958}. The observable consequences of this phenomenon in electronic systems were later uncovered~\cite{Patrick1985,mott2012electronic}, and its direct experimental observation was achieved for photons in disordered lattices and photonic crystals~\cite{wiersma1997localization,schwartz2007transport,Lahini2008}, and also for acoustic waves~\cite{weaver1990anderson,hu2008localization} and other types of waves in condensed matter~\cite{billy2008direct}. In the context of thermal phonon transport, early works identified the significance of localization in systems with crystal lattice disorder~\cite{Lebowitz1974,Allen1993}. From a mechanistic perspective, coherent phonon evolution via harmonic interactions within disordered environments was predicted to unlock analogous stochastic interference patterns and localization~\cite{allen1998evolution,allen1999diffusons}. More recently, the role of Anderson localization (AL) in heat conduction in the presence of anharmonic effects has been investigated under a variety of conditions~\cite{Savic2008,YanWang2014, luckyanova2018phonon,hu2018randomness,juntunen2019anderson,RenjiuHu2021}. Furthermore, integration of this kind of wave-like behavior with the usual particle-like phonon description has been shown critical to predicting the thermal conductivity in amorphous materials~\cite{Simoncelli2019,isaeva2019modeling}. Given its nondeterministic nature, precise spatial and temporal manipulation of AL in a predetermined manner is challenging and fundamentally limited to relatively high frequencies due to the phonon interference limit, whether the random perturbation is at the primitive-cell~\cite{Savic2008} or artificial-cell~\cite{YanWang2014, luckyanova2018phonon,hu2018randomness,juntunen2019anderson,RenjiuHu2021} level.

In ordered crystalline materials, realization of any direct and explicitly observable form of atomic-scale phonon localization has remained elusive.~First, similar to AL, wave-like behavior of phonons is required at length and time scales smaller than at least a portion of the mean free path (MFP) and lifetime distributions, respectively. Under these conditions, coherent behavior may be realized by Bragg scattering of phonons due to the presence of periodic nanoscale structural features such as repeated layers in a superlattice (SL)~\cite{barker1978study,yamamoto1994coherent,venkatasubramanian2000lattice,chen2005minimum,mcgaughey2006phonon,simkin2000minimum,ravichandran2014crossover} or an array of holes or inclusions in a nanophononic crystal (NPC)~\cite{tang2010holey,yu2010reduction,davis2011thermal,he2011thermal,zen2014engineering,Minnich2014,yang2014extreme,feng2016ultra,lee2017investigation,hu2018randomness}.~Band flattening causes reduced group velocities, which leads to inhibited transport but not distinct spatial phonon localization. Some mild level of localization was reported in  SLs~\cite{venkatasubramanian2000lattice} and NPCs~\cite{he2011thermal,yang2014extreme,feng2016ultra}, although in environments where incoherent scattering effects cause behavior that may appear as localization$-$often with difficulty in making a clear distinction with coherent response.~A key challenge is that nanoscale Bragg scattering requires almost defect-free conditions and near-perfect internal surfaces~\cite{lee2007lattice,Minnich2014,lee2017investigation}. In addition to band flattening,  phonon band gaps may in principle appear in SLs~\cite{chen2005minimum,mcgaughey2006phonon} and NPCs~\cite{zen2014engineering} especially at low temperatures. However, the complex nature of the atomic-scale dispersion branches usually limits the possibility of full band gaps particularly in multi-dimensional arrays~\cite{davis2011thermal}. \\
\indent Distinct from Bragg scattering, a recently discovered mechanism that has been shown to cause transformative changes to the phonon band structure is engineered intrinsic atomic-scale local resonances~\cite{Hussein2014}~\footnote{The notion of intrinsic local resonances has been introduced in the context of electromagnetics~\cite{pendry1999magnetism,smith2000composite} and acoustics~\cite{liu2000locally,pennec2008low,wu2008evidence}.~Unlike thermal phonons, these applications are practically described as linear wave problems where the physical behaviour of interest is dominated by excitations at a single frequency or a narrowband frequency range.}.~The presence of sub-MFP vibrating substructures generates a multitude of wave-vector-independent resonant phonons (also described as vibrons~\cite{Hussein2018Handbook}) that couple and hybridize with the underlying traveling phonons in a host medium~\cite{hussein2020thermal}. Nanostructures exhibiting this behavior are referred to as \textit{nanophononic metamaterials} (NPMs); a wide range of NPM configurations have been investigated~\cite{wei2015phonon,honarvar2016spectral,xiong2016blocking,iskandar2016modification,honarvar2018two,giri2018giant,ma2019quantifying,neogi2020anisotropic,wang2021synergistic,wang2021anomalous,li2022phonon,chen2022phonon,liu2023enhanced,anufriev2023impact,li2024phonon}.~Unlike the Bragg scatterers in an NPC, the nanoresonators in an NPM are not necessarily periodically spaced, which enhances the robustness of the mechanism.~Thermal conductivity reduction by nanoscale local resonances has recently been experimentally demonstrated in suspended silicon membranes with a random array of GaN nanopillars standing on the surface~\cite{spann2023semiconductor}. This nanopillars-on-a-membrane configuration also yielded a decoupling between the thermal and electrical properties, which is highly advantageous for thermolectrics~\cite{chen2003recent,vineis2010nanostructured}.\\
\indent Here we reveal$-$for the first time$-$the appearance of~\textit{spatial phonon localization along the direction of transport} in this class of nanostructures.~We provide evidence of this phenomenon by direct explicit observations from room-temperature atomic-scale simulations of silicon membranes with resonating silicon nanopillars standing on the surface.~These  simulations involve frequency-targeted wave-packet excitations that are applied over a thermalized atomic environment at equilibrium, where the full scope of phonon-phonon scattering events is accounted for~\footnote{This contrasts sharply with the localization of linear elastic waves due to local resonances at the macroscopic scale~\cite{oudich2011experimental}.}.~The form of phonon localization we highlight, which we refer to as \textit{resonant localization}, occurs in the base membrane along the in-plane direction across multiple unit cells.~This is related but distinctly different from the observation of a large 
concentration of phonon motion along the orthogonal direction in a nanopillar in a single unit cell as described by mode shapes obtained by harmonic eigenvalue analysis$-$with the latter often also described as a form of localization in the literature~\cite{wei2015phonon,honarvar2018two,hussein2020thermal}. A localized eigen mode is associated with a specific frequency and wavevector pair, whereas the resonant localization that we demonstrate is associated with a single frequency or a narrow band of frequencies and the full scope of all possible wavenumbers along the direction of transport. Resonant localization has profound impact on the thermal behavior  and other condensed matter properties in general, and its assessment along the direction of transport enables direct comparison with AL and with coherent phenomena in other types of ordered nanostructures.~For example, similar to AL, the localization length or the fraction of localized energy may be characterized.~(1) Unlike AL and other coherent phonon mechanisms, this distinct phenomenon manifests across all the phonon spectrum, including low frequencies smaller than the mobility edge of a randomly perturbed lattice (AL)~\cite{john1983localization,Sheng2006Book,luckyanova2018phonon} or the Bragg limit corresponding to the periodicity of ordered nanostructure features (SL and NPCs)~\cite{hussein2020thermal}. (2) Furthermore, unlike AL it is fully deterministically predictable in both space and time.~(3) Another unique aspect of the resonant localization mechanism is that it is associated with minimal energy exchange with external sources at the resonant frequency bands. \\
\indent We first characterize the anharmonic phonon band structure of an NPM to highlight the unique influence of the nanoresonators at room temperature.~We then run the simulations where narrowband wave-packet excitations~\cite{Tlili2019} are applied to thermalized molecular dynamics (MD) models of the NPM. We aim our analysis to allow for an elucidation of the propagation and external energy absorption characteristics of phonons within resonant frequency ranges, and, in contrast, within non-resonant frequency ranges.~These controlled excitations under fully anharmonic conditions explicitly expose the occurrence of the resonant localization phenomenon, which was not possible in earlier studies limited to either a single-unit-cell eigenvalue analysis under the harmonic approximation ~\cite{wei2015phonon,honarvar2018two,hussein2020thermal} or standard nonequilibrium MD simulations of multiple unit cells in the steady state~\cite{zhang2023surface}.~We also run the same simulations on uniform (unpillared) membranes with intentionally introduced atomic-scale disorder to provide a direct comparison between resonant localization and Anderson localization, demonstrating the fundamental differences between the two mechanisms. 

\begin{figure}
    \centering
    \includegraphics{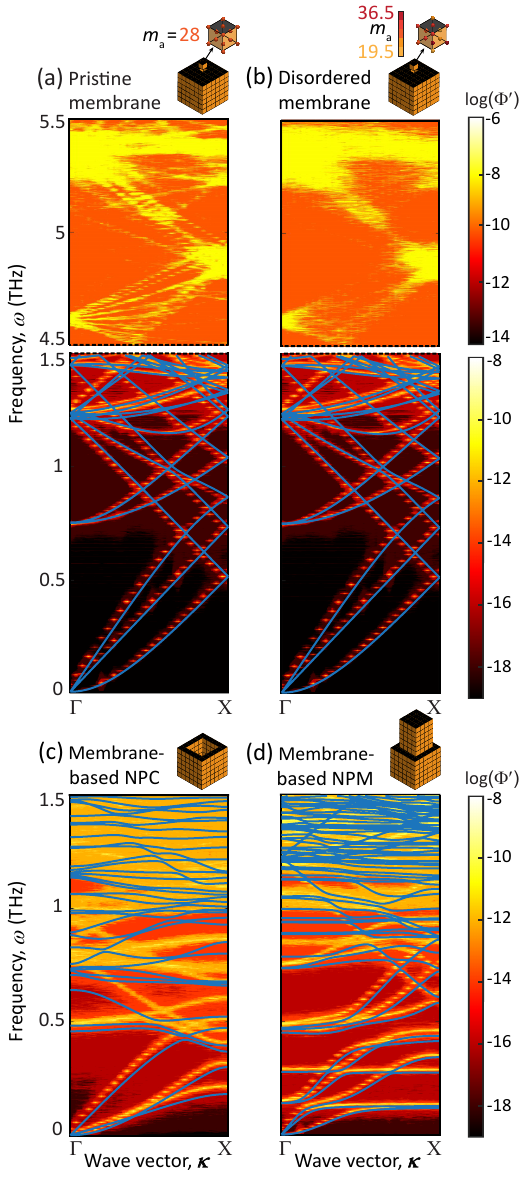}
    \caption{\textbf{Phonon dispersion band structure from harmonic LD calculations (solid lines) and anharmonic MD simulations followed by SED calculations (color) for different silicon nanostructures.} The geometry schematics represent a periodic unit cell in each case. The thickness of the systems (a), (b), (c) and the base membrane in (d) is 3.26 nm. The atomic mass $m_{\text{a}}$ in (b) is randomly varied around the nominal Si atom mass $m$ according to a Gaussian distribution. More details on the geometries considered and the methods to compute the band structures are provided in Appendix \ref{methods_geometries} and \ref{methods_dispersion}, respectively.}
    \label{fig:Fig1}
\end{figure}

\section{Phonon Band Structures}\label{bandstructure}
Nanostructuring influences the nature of coherent phonon motion when the MFP is large compared to the length scale of the nanostructure features at the temperature of interest. For comparison, we consider four configurations of suspended ultrathin silicon membranes, namely, a pristine membrane, an atomically disordered membrane, a membrane with an array of nanoholes (i.e., NPC), and a membrane with an array of nanopillars (i.e., NPM). Unlike NPCs, periodicity is not necessary for an NPM; however it is assumed here to allow for phonon band structure calculation. The thickness of the membrane for the first three cases and of the base membrane in the NPM is 3.26 nm. The unit-cell size along each of the two planar directions is also 3.26 nm for all configurations.~All other information on the unit-cell geometries, disorder parameters (for the second case), and feature sizes (for the third and fourth cases) are given in Appendix~\ref{methods_geometries}, and images of the unit cells are depicted in Fig.~\ref{fig:Fig1}.~Using the Stillinger-Weber interatomic potential \cite{SWpotential}, we run room-temperature equilibrium MD simulations followed by spectral energy density (SED) calculations~\cite{McGaughey2010,honarvar2016spectral,hussein2020thermal} on each system to obtain the corresponding anharmonic phonon band structures, which are also shown in Fig.~\ref{fig:Fig1}. The conventional lattice dynamics (LD) dispersion curves based on the harmonic approximation are overlaid in the figure for each case for direct comparison. Modeling and computational details are provided in Appendix~\ref{methods_geometries} and~\ref{methods_dispersion}, respectively. 


We begin by analyzing the influence of disorder, which is introduced by randomly varying the mass of each silicon atom around the nominal value. The phonon dispersion curves of a pristine (ordered) membrane and a corresponding disordered membrane are shown in Figs.~\ref{fig:Fig1}a and~\ref{fig:Fig1}b, respectively.~The wave-like properties of low-frequency phonons, at $\omega<1.5$ THz, are practically identical in the two cases. However, at higher frequencies, $\omega>4.5$ THz, the dispersion band structure becomes almost indistinguishable in the presence of disorder.~This is due to the loss of Bloch-wave coherence of the phonons, and its replacement with stochastic coherent scattering due to the introduced mass disorder, which hinder dispersive wave-vector-dependent phonon propagation through the medium~\cite{Lebowitz1974,allen1998evolution}.~This behavior is a manifestation of AL, where phonon propagation at frequencies higher than the mobility edge at the primitive unit-cell level is partially impeded causing a reduction in the lattice thermal conductivity. \\
\indent We next examine the phonon band structure of an NPC unit cell, comprising a membrane with a central squared nanohole as shown in Fig.~\ref{fig:Fig1}c.~This is to demonstrate the difficulty of realizing phonon attenuation under anharmonic conditions in this class of nanostructures.~Here we observe significant flattening of the bands across the spectrum compared to those of the uniform membrane of Fig.~\ref{fig:Fig1}a; this is due to phonon interferences based on Bragg scattering created by the presence of the periodically arranged nanoholes.~According to the harmonic LD calculations, this effect is distinctly observable at relatively low frequencies and, in fact, opens a band gap centered around~$\omega\sim$~0.7 THz$-$thus indicating that phonon propagation within this specific frequency range is attenuated. However, in contrast, the anharmonic MD-SED contour shows band shifts leading to a complete closure of this band gap with all the energy within the frequency range of the harmonic band gap now being fully accommodated into well-defined propagating anharmonic phonon modes.~This illustrates that, even though the emergence of band gaps might be apparent in models approximated by harmonic atomic interactions, the influence of anharmonicities may prevent their prevalence under realistic phonon-phonon scattering conditions. Therefore, an NPC, even in the ideal case of no imperfections, is generally not a nanosystem capable of robustly sustaining the presence of phonon band gaps and the phonon attenuation they may cause. \\
\indent Moving to NPMs, Fig.~\ref{fig:Fig1}d shows the anharmonic phonon band structure for a resonant unit cell in the form of a membrane with a central squared nanopillar having the same height as the membrane.~As seen in previous studies~\cite{honarvar2016spectral,hussein2020thermal}, both the LD and MD-SED calculations show in the dispersion diagram the hybridization of numerous phonon modes in this geometry, distinguishable as horizontal lines associated with resonances coupled with the underlying host medium's wave-vector-dependent propagating modes.~The mode shapes of eigenstates populating these horizontal lines (and parts of the associated avoided crossing regions) indicate a concentration of the atomic kinetic energy in the nanoresonator portion of the unit cell, e.g., inside the nanopillar portion of a nanopillared membrane~\cite{wei2015phonon,honarvar2018two,hussein2020thermal,zhang2023surface}.~In contrast to the loss of band gaps in NPCs, band shifts stemming from anharmonicity do not disrupt the presence of the local resonances and their ability to couple with propagating waves, even at room temperature$-$as shown for the silicon NPM of Fig.~\ref{fig:Fig1}d.~The occurrence of the local resonances may in principle be precisely synthesized to take place across the full phonon spectrum.~Indeed, it is also observed in Fig.~\ref{fig:Fig1}d that the resonance hybridizations take place across the full frequency range displayed, including at low frequencies below the Bragg limit associated with the size of the NPM unit cell.~In principle, the number and specific frequencies of these resonances are a characteristic of the nanopillar geometry and choice of material, and are therefore controllable by design. The number of local resonances is three times the number of atoms in a unit nanoresonator, and the bandwidth of each resonant mode is dependent on its dissipation properties~\cite{honarvar2018two}.~Anharmonic unit-cell analysis may also be done by lattice dynamics-based calculations; for example the quasiharmonic Green-Kubo method~\cite{isaeva2019modeling} was used to predict a frequency-dependent diffusivity quantity corresponding to the anharmonic phonon band-structure of an NPM unit cell~\cite{neogi2020anisotropic}.~\\
\indent Distinct isolation of the resonance hybridization effects in standard MD simulations or in experiments is challenging because the thermal conductivity is influenced by the collective behavior of all resonant modes$-$thus the response of a particular resonance or a set of closely grouped resonance modes is unlikely to be explicitly accessed, and to the authors' knowledge has not been isolated and exposed in any atomic-scale investigation in the literature. More broadly, mechanistic classification is further complicated because the thermal conductivity in general is also influenced by other mechanisms at the nanoscale such as incoherent reduction of the phonon lifetimes and diffusive scattering from internal defects and non-perfect surfaces.~In the following sections, we investigate the implications of the band signatures observed in Fig.~\ref{fig:Fig1} by directly simulating the propagation of specific phonon modes within a selection of narrowband frequency ranges under anharmonic conditions.~This approach enables the elucidation of deterministic localization of thermal phonons via local resonances, and, as we show, in perfect consistency with the anharmonic dispersion band structure. 

\section{Deterministic localization via local resonances} \label{deterministic}

\begin{figure*} [t!]
    \includegraphics{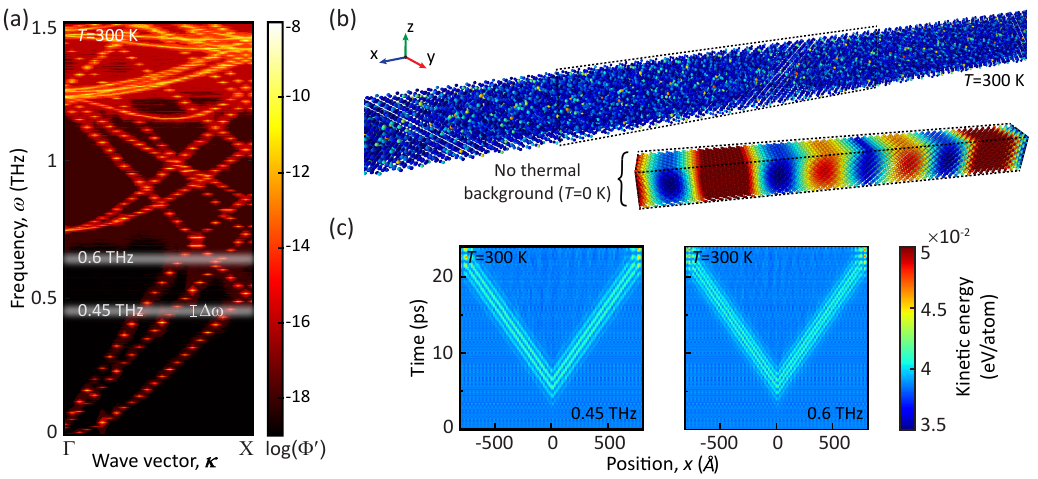}
    \caption{\textbf{Wave-packet propagation in pristine membrane.} (a) MD-SED phonon band structure. (b) Phonon kinetic energy in each atom at a representative time-step after excitation at a frequency of $\omega$=0.45 THz. The inset shows the non-equilibrium energy distribution with the the thermal background removed, i.e., $T=0$ K response. (c) Phonon kinetic energy density distribution in space and time at $T=300$ K after excitation at different frequencies.}
    \label{fig:Fig2}
\end{figure*} 
\begin{figure*} [t!]
    \includegraphics{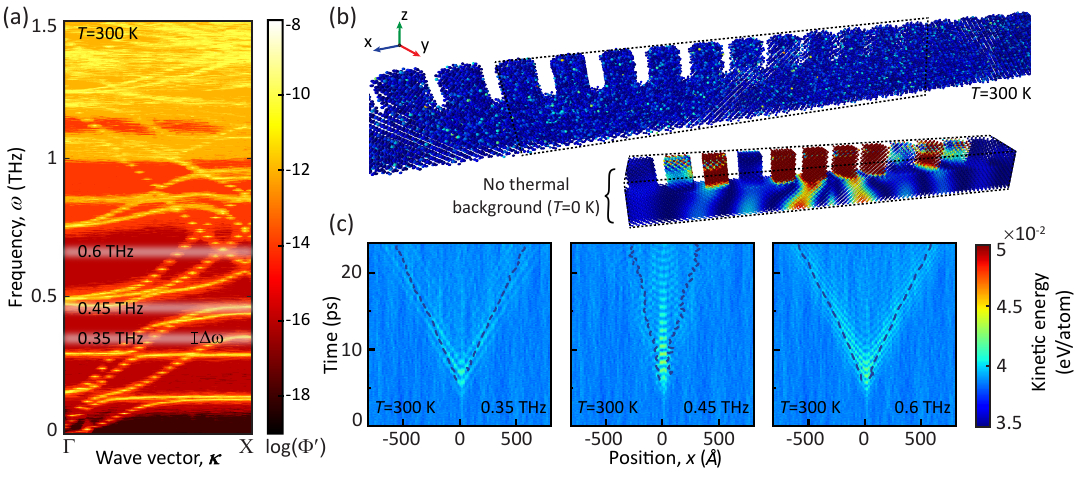}
    \caption{\textbf{Wave-packet propagation and localization in NPM.} (a) MD-SED phonon band structure. (b) Phonon kinetic energy in each atom at a representative time-step after excitation at the resonant frequency of $\omega$=0.45 THz. The inset shows the non-equilibrium energy distribution with the thermal background background removed, i.e., $T=0$ K response. (c) Phonon kinetic energy density distribution in space and time at $T=300$ K after excitation at different frequencies. The region containing 70$\%$ of the wave-packet energy is bounded by dashed lines. At the resonant frequency $\omega=0.45$ THz, a significant amount of energy is  locally restrained to the excitation region, demonstrating the phenomenon of resonant localization.}
    \label{fig:Fig3}
\end{figure*}
In this section, we establish the occurrence of the resonant phonon localization phenomenon.  We consider the same silicon membrane and NPM models introduced in Fig.~\ref{fig:Fig1}.~We use the wave-packet simulation methodology~\cite{Tlili2019} on models based on the same anharmonic interatomic potential~\cite{SWpotential} considered for the MD-SED simulations of Section~\ref{bandstructure}. This approach builds upon previous research on the wave nature of phonons \cite{allen1998evolution}, and has recently been shown useful for the study of non-diffusive transport effects \cite{Desmarchelier2021,Desmarchelier2021a,desmarchelier2024phonon}.~First, we run room-temperature equilibrium MD simulations on an extended model of the structure until it is fully thermalized. Then, a wave-packet is excited by perturbing the atoms within a 4 \AA-wide vertical layer located at the center of the model using a force in the form 
\begin{equation}
f=A \sin[ 2 \pi\omega^*(t-3\tau)]\exp{\left[-\frac{(t-3\tau)^2}{(2\tau^2)}\right]},
\label{eq:forceeq}
\end{equation} 
where $\omega^*$ is the central frequency of excitation, $A$=0.1 eV/\AA\ is the amplitude, $\tau$ = 2 ps is the excitation time-window, and $t$ denotes time.~The force is polarized in the out-of-plane transverse direction relative to the membrane plane and the wave-packet propagation$-$generating shear motion.~After applying the excitation, the kinetic energy distribution throughout the computational domain is monitored to allow us to study the phonon propagation behavior for the given excitation frequency. To resolve the non-equilibrium energy from the thermal background, the results are averaged over 10 independently run simulations with different initial conditions. More details on the computational domains and simulation procedure are provided in Appendix~\ref{methods_WP}.
\begin{figure*}
    \centering
    \includegraphics{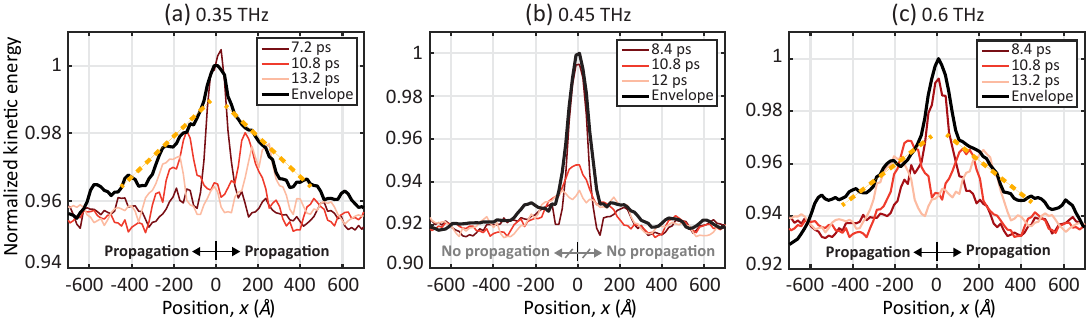}
    \caption{\textbf{Wave-packet envelope evolution at room temperature.} Maximum phonon kinetic energy measured around the intervention region resulting from wave-packet excitation at the resonant frequency (b) or non-resonant frequencies (a,c) in NPM. The kinetic energy is represented in logarithmic scale, and the dashed lines indicate exponential fits to the envelope relaxation. As shown in (b), this fit is unsuccessful under the resonant conditions due to the lack of wave front propagation, thus indicating unusual phonon propagation and relaxation beyond the particle-like scattering picture provided by the relaxation time approximation.}
    \label{fig:Fig4}
\end{figure*}\\
\indent In Figs.~\ref{fig:Fig2} and \ref{fig:Fig3}, we consider wave-packet excitations at different frequencies and show the propagation of phonon motion, represented by the kinetic energy, along the membrane without and with the nanopillars, respectively. The thermal background prevents visual identification of the wave-packet energy in 3D atomistic representations (Figs.~\ref{fig:Fig2}b and \ref{fig:Fig3}b). However, cross-plane integration of the kinetic energy at a sampling of time instants enables resolving the propagation characteristics, as shown in Figs.~\ref{fig:Fig2}c and \ref{fig:Fig3}c. In the absence of nanopillars, for all excitation frequencies, the wave-packet propagates at a well-defined velocity in good agreement with the group velocity d$\omega$/d$\kappa$ determined by the anharmonic dispersion relation at each frequency, where $\kappa$ is the wavenumber along the $\Gamma$X direction. For example, at 0.6 THz, the wave-packet propagation velocity estimated from the kinetic energy map is 4.88 nm/ps, and the slope of the shear-polarized band is d$\omega$/d$\kappa=4.78$ nm/ps. In contrast, we detect fundamentally distinct propagation behavior in the NPM depending on the excitation frequency. We consider, for example, the resonant frequency $\omega=0.45$ THz identified in the vicinity of a set of horizontal lines representing local resonances in the anharmonic phonon band structure of Fig.~\ref{fig:Fig3}a. On the one hand, for excitation frequencies  below or above this resonant frequency, we obtain wave-packed propagation with a well-defined velocity as observed in the unpillared membrane. On the other hand, the wave-packet excited at the resonant frequency is not able to propagate and experiences intense and highly distinguishable phonon localization behavior, as shown in Fig.~\ref{fig:Fig3}c. Finally, we note that the wave-packet velocity for the NPM at the non-resonant  frequency of 0.6 THz is smaller than in the standard membrane. This is a consequence of a degree of flattening of the dispersion curves even outside the hybridization regions due to the periodic placement of the nanopillars, which brings rise to some Bragg scattering behavior over and above the resonant behavior. \\ 
\indent The emergence of this unique phonon localization phenomenon is identifiable across the full phonon spectrum.~As shown in Appendix \ref{localized_energy_fraction} in terms of the fraction of spatially localized energy with respect to the total energy, each mode hybridization detected in the anharmonic phonon band structure induces analogous localization effects in the wave-packet simulations.~In particular, we observe resonant localization even at very low frequencies in the sub-Bragg regime (i.e., for first Brillouin-zone phonons with wavelengths larger than the periodic unit-cell size of the NPM as well as short wavelength phonons in higher Brillouin zones that correspond to the same low frequencies). Therefore, in contrast to other mechanisms such as AL and Bragg scattering, the characteristic scale of the NPM system does not set a lower limit to the phonon frequency range prone to localization. Similar results are obtained if considering a force polarized in the longitudinal or flexural directions (see Appendix \ref{localized_energy_fraction}). Since the resonant hybridization strength is not equivalent for different modes at a given resonant frequency, the fraction of localized energy is dependent on the excitation polarization. \\
\indent These observations imply strong coupling between the modes in the membrane and the standing waves in the nanopillars at some specific frequencies.~For illustration, in Figs.~\ref{fig:Fig2} and \ref{fig:Fig3} we also show the kinetic energy distribution after excitation by removing the thermal background. Since we use a resonant excitation frequency in the NPM, it is observed that most of the energy is stored in the nanopillars, and the energy in the supporting base membrane becomes partially localized below them. Further discussion on the coherent interferences of phonons from different regions of the NPM system is provided in Appendix~\ref{memb-pillar_coupling} by analyzing the SED spectrum considering atoms in different portions of the nanopillared membrane unit cell and by dissecting the spatial distribution of the wave-packet kinetic energy across the unit cell geometry. Most importantly, both the localization and group velocity reduction of specific modes hinders efficient heat conduction, thus indicating an increase in the time scales for thermal relaxation in NPMs. For the particular geometries and sizes considered here, we obtained a 3.6-fold reduction of the in-plane thermal conductivity in the NPM relative to the unpillared membrane using the Green-Kubo (G-K) method (see Appendix~\ref{methods_Green-Kubo} for details). \\
\indent To further explore the resonant localization phenomenon, it is illustrative to analyze the envelope of the wave-packet energy as a function of the position along the NPM structure in the direction of transport, as shown in Fig.~\ref{fig:Fig4}.~At non-resonant frequencies, the amplitude of the wave packet exponentially decays in space. This attenuation is a signature of the anharmonic interaction of the excited phonons with the rest of the phonon population and the boundaries.~Nonlinear transfer of wave-packet energy to other phonon modes dissipates the energy accommodated in the wave front into the thermal background, thus leading to diffusive evolution at long times.~Precisely, the characteristic length scale of the decay multiplied by the phonon group velocity can be interpreted as the phonon mode relaxation time due to phonon scattering.~This interpretation is confirmed by wave-packet simulations at 0 K, where the lack of anharmonic effects prevents the decay of the wave-packet amplitude.~A key observation is that an exponential relaxation of the wave-packed is similarly not observed when a resonant frequency is excited in the NPM at 300 K. In this case, the excited phonon modes are fully localized, and the only available mechanism to relax the perturbation is the nonlinear transfer of energy to modes with non-resonant frequencies that can propagate. In Figure Fig.~\ref{fig:Fig4}b, this dissipative relaxation process translates to both a lack of wave fronts and a non-exponential decay of the wave-packet envelope. These effects are thus unambiguous signatures of deterministic localization that can be qualitatively distinguished from velocity reduction effects due to band flattening. In turn, these results involving resonances indicate that the particle-picture of independent phonon evolution based on a velocity and a characteristic anharmonic relaxation time for each mode is insufficient, and highlights the need of integrating the wave and particle pictures of phonons  \cite{Simoncelli2019,isaeva2019modeling,ZhangPRL2022} to develop mesoscopic descriptions of heat transport in resonant metamaterials.\\
\indent We recall that the wave-packet simulations are performed within a thermodynamic environment at 300 K as described by a fully anharmonic interatomic potential accounting for phonon scattering and associated entropy generation~\cite{beardo2024entropic}.~However, the simulated wave packet is a prescribed mechanical disturbance that selectively excites a frequency range instead of a thermal excitation that perturbs a wide range of phonon modes.~This results in an artificial phase-space for phonon interaction where the majority of modes remain close to equilibrium.~Moreover, our simulations assume idealized nanostructures with atomically flat surfaces, which promote specular phonon-boundary scattering and enhance the coherent behavior of phonons compared to experimental conditions in samples presenting defects and roughness. Nevertheless, these results explicitly demonstrate that the emergence of phonon mode hybridizations in the band structure fundamentally modifies the propagation of thermal phonons even at room temperature.~These results will further guide experimental research, noting that there are cases where the impact of the resonances has been observed, e.g., on the thermal conductivity~\cite{spann2023semiconductor} and on the low-frequency band structure~\cite{anufriev2023impact}, but also cases where such characterization was not possible~\cite{huang2019thermal,huang2020coherent,lees2023nanoscale}\footnote{Among the experimental challenges for NPM characterization are: (1) the presence of fabrication defects or imperfections, which enhance the role of anharmonicity at room temperature relative to atomically-perfect crystals, (2) fabrication of prototypes with a very low pillar-to-membrane volume ratio, which weakens the impact of the local resonances, and (3) examination of configurations where the unit cell is overly large and comparable to the mean free path, which weakens the degree of presence of coherent phonons.} At lower temperatures, the resonant localization effect persists even more strongly as discussed in Appendix~\ref{low_temperatures}. 

\section{Minimization of energy exchange with external sources in resonant conditions}\label{work_min} 

The wave-like effects on phonon transport created by the resonating substructures are analogous to macroscale behavior in acoustic and elastic metamaterials in the sub-GHz regime~\cite{liu2000locally,hussein2014dynamics,jin2021physics}.~This analogy permits extrapolating known results from macroscale elastodynamics to thermal phonon transport.~One important aspect of phononic structures at the macroscale is that the energy absorbed by an external force depends on the extent a resonance or an antiresonance is excited, see, e.g.,~\cite{kianfar2023phononic}. In this section, we demonstrate that this behavior also takes place when considering frequency excitations in the THz regime in relation to thermal phonons, and in particular to NPMs given their unique resonance properties. \\
\indent Consider the work $W$ performed by the external excitation $f$ (Eq.~\eqref{eq:forceeq}), or, equivalently, the energy increase in the system during the excitation time-window:
\begin{equation}
W_f=\sum_i^n\int_{\Gamma_i} f\textbf{e}\cdot \text{d}\textbf{s},
\label{eq:work}
\end{equation} 
\begin{equation}
W_e=\sum_i^N \bigg(E_i(t=10\tau)-E_i(t=0)\bigg),
\label{eq:work2}
\end{equation} 
where the sums run over the $N$ atoms in the system, or the $n$ atoms within the 4 $\AA$-long excitation region, respectively, $E_i$ is the kinetic energy of atom $i$ at a given time $t$, $\textbf{e}$ is the unitary vector parallel to the force polarization, and $\text{d}\textbf{s}$ is the differential displacement vector along the atomic trajectory $\Gamma_i$ during the excitation. Since the perturbation takes place without thermostats nor barostats connected to the system, the excitation force $f$ is the only source of energy. Consistently, both methods to obtain the work $W$ done on the nanostructure from the MD simulations provide very similar numerical results, i.e., $W\approx W_f\approx W_e$.~Here we use the energy increase calculation after verifying that $W_e$ minimizes the numerical error (see more details in Appendix \ref{methods_WP}). \\
\indent In Fig.~\ref{fig:Fig5}, we show that the work $W$ performed on the NPM system from an external source depends non-monotonously on the frequency $\omega$ and displays minima around resonant frequencies, where mode hybridizations are identified in the anharmonic phonon band structure (see Fig.~\ref{fig:Fig1}d).~Furthermore, an excitation frequency that minimizes the external work also maximizes the fraction of localized energy within the NPM (see Appendix \ref{localized_energy_fraction}). This implies that the energy accommodation among the different modes is perturbed by the presence of local resonances, not only for the present idealized excitation, but also under realistic thermal perturbations, where multiple frequencies are simultaneously excited. By contrast, the work performed on the membrane without nanopillars displays a monotonic trend with the excitation frequency. Overall, these results imply that not only the energy transfer within the metamaterial can be manipulated by nanostructured resonating substructures, but also the amount of thermal energy absorbed from external sources can be influenced.\\
\begin{figure} [t!]
    \includegraphics{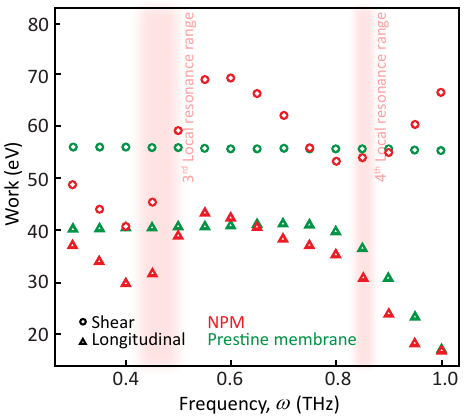}
    \caption{\textbf{External work as a function of the wave-packet excitation frequency} in membrane (green) and NPM (red) nanostructures. Both shear (circles) and longitudinal (triangles)  excitation polarizations are considered. The energy exchange with the external source is minimized under resonant conditions.}
    \label{fig:Fig5}
\end{figure}
\begin{figure*}
    \includegraphics{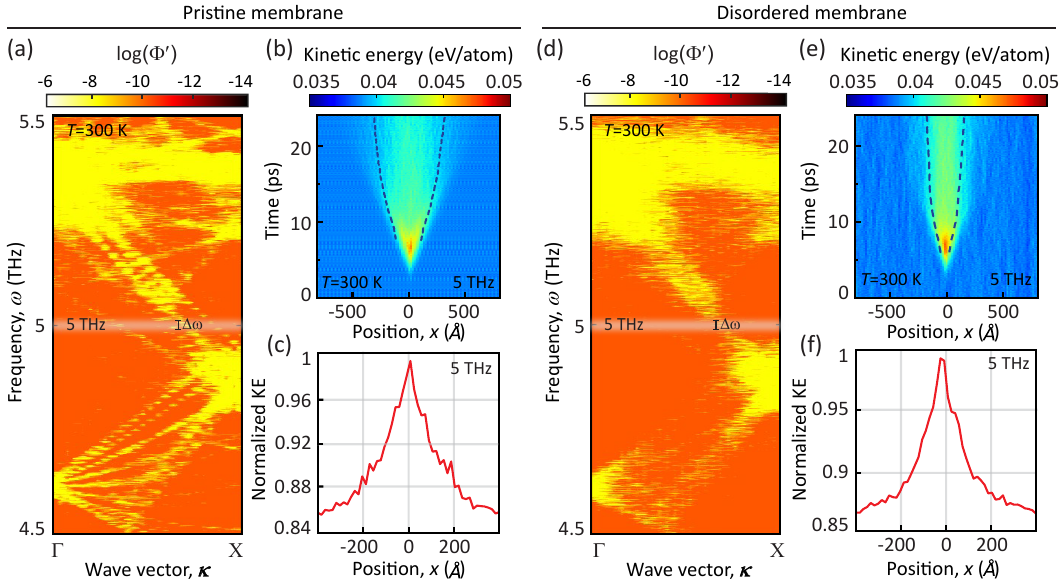}
    \caption{\textbf{Anderson localization induced by atomic mass disorder at the crystal's primitive cell level.} The MD-SED band structure (a,d), and the wave-packet spatiotemporal evolution (b,e) and envelope (c,f) are shown for pristine and mass-disordered membranes, respectively. The border of the region containing 70$\%$ of the wave-packed energy is identified by dashed lines in plots (b,e).}
    \label{fig:Fig6}
\end{figure*}
\indent To further demonstrate the robust connection between the NPM's resonant phonon band structure and the phonon propagation mechanisms, in Fig.~\ref{fig:Fig5} we also include the work performed by a longitudinally polarized external force. In the NPM anharmonic phonon band structure (Fig.~\ref{fig:Fig1}d), we observe that around 0.45 THz hybridization of the modes take place for all the bands. In contrast, not all the bands present mode hybridizations at higher frequencies around 0.85 THz.~Consistently, we observe that the minimization of the work appears for both shear and longitudinal polarizations around the low-frequency resonance, while only for the shear polarization at the higher frequency$-$a minimum is absent for the longitudinal excitation. This indicates that the dominant modes coupling the base membrane and nanopillar dynamics around 0.85 THz predominantly display shear polarization with weak or no coupling of longitudinal motions. Finally, in Fig.~\ref{fig:Fig5}, we also observe that, for longitudinal excitation, the work monotonously decreases for $\omega>0.8$ THz in both the membrane and the NPM systems. We attribute this to the extremely low group velocity of the corresponding modes, which bottlenecks the amount of energy that can be injected during the excitation time window, thus limiting the work applied in these conditions. \\
\indent The correlation of the work minimization and the mode hybridizations at certain frequencies demonstrate that the introduction of resonant nanostructures might not only be useful to influence the thermal conductivity for thermoelectric applications but also for transient manipulation of thermal energy transfer between the NPM and the environment.  These results might hint to new ways to engineering energy pathways in semiconductor nanosystems with a variety of potential applications including phononic computing~\cite{Baowen2012}. Furthermore, the results in Fig.~\ref{fig:Fig5} may inspire control of the total energy absorption from laser sources~\cite{yoon2024terahertz} and contact nanotips~\cite{zhang2020review}, thus motivating further new experiments on the atomic-scale local resonances phenomenon.

\section{Anderson localization via 
\\
mass disorder}\label{stochastic}

Now we proceed to utilize the wave-packet simulation method to examine AL to allow for a comparison with resonant localization within a common analysis framework.~We model AL by introducing disorder at the crystal's primitive-cell level, i.e., without the creation of an artificial unit cell.~We consider the pristine membrane and mass-disordered membrane systems introduced in Fig.~\ref{fig:Fig1}.~As detailed in Appendix \ref{methods_geometries}, the disorder is realized by randomly varying the mass of the silicon atoms according to a Gaussian distribution with 10$\%$ relative standard deviation around the nominal atomic mass.~The resulting phonon coherent and incoherent scattering due to these mass defects mimic the role of vacancies, isotopes, impurities, or alloying, in realistic samples, and are expected to generate stochastically-varying phonon interference patterns that can lead to localization \cite{Savic2008,Howie2014,zhang2023phonon}.~To excite the lattices, we implement the same wave-packet simulation protocol described in Section \ref{deterministic} and Appendix \ref{methods_WP} using the same parameter values in Eq.~\eqref{eq:forceeq} and considering shear polarization. \\
\indent As described earlier, the AL mechanism is limited to frequencies that are higher than the mobility edge, which may be approximated by examining the interference limit in the acoustic branches of the dispersion curves of the underlying unperturbed crystal lattice or, alternatively, the average lattice spacing of the random scattering centers.~Furthermore, according to Rayleigh's law, the rate of elastic scattering of phonons with point defects is proportional to the fourth power of the phonon frequency $\omega$~\cite{klemens1955scattering}.~Hence, the effect of defects generally weakens as the frequency decreases.~Consistent with these limitations, for low excitation frequencies, $\omega<1.5$ THz, our wave-packet simulations display propagation features in the disordered membrane that are almost identical to the results shown in Fig.~\ref{fig:Fig2} for the pristine membrane (see quantitative comparison in Appendix~\ref{localized_energy_fraction}). This is in line with the observation that there is no influence of disorder in the MD-SED dispersion band structure below 1.5 THz in Fig.~\ref{fig:Fig1}.\\
\indent In contrast, at higher frequencies, where the phonon wavelengths shrink, we observe a distinctly different wave-packet response between the two systems, as shown in Fig.~\ref{fig:Fig6} for $\omega=5$ THz.~In the pristine case, a significant amount of the phonon energy propagates through the lattice, whereas in the disordered case most of the energy remains localized in the excited region.~In spite of this behavior being a signature of wave-like harmonic evolution in the presence of disorder \cite{allen1998evolution,Sheng2006Book}, Fig.~\ref{fig:Fig6} quantitatively exposes its impact on wave-packet propagation in fully thermalized environments where anharmonic effects are accounted for.~We note that in our perturbed model, no correlations are present in the disorder pattern. In the literature, long-range correlations have been applied to disorder and shown to influence the properties of AL including the localization length~\cite{Sajeev1983,Chu1989}\\
\indent The localization of phonons by AL reduces the lattice thermal conductivity of the material~\cite{Savic2008,YanWang2014, luckyanova2018phonon,hu2018randomness,juntunen2019anderson,RenjiuHu2021}$-$a matter of practical importance. Relative to the pristine membrane, we obtain a 1.8-fold reduction in the thermal conductivity in the disordered case as evaluated using the G-K method following equilibrium MD simulations (see Appendix \ref{methods_Green-Kubo}). However, the thermal conductivity reduction might also be influenced by diffusive scattering due to the disordered atoms also acting as defects, a phenomenon present even in the glass limit \cite{Simoncelli2019,isaeva2019modeling}~\footnote{Beyond AL, incorporation of long-range spatial correlations in the defect distribution has recently been shown to enhance the interaction between defects and acoustic phonons, with an effect on the thermal conductivity due to life-time reductions~\cite{Thebaud2023}.}. Overall, the effect of disorder on the thermal conductivity is shown to be significantly weaker than resonances in the present system realizations with comparable length scales.\\
\indent By inducing atomic-scale disorder at the length scale of an artifical unit cell instead of the interatomic distance of the crystal's primitive unit cell, longer wavelength acoustic phonons become prone to localization. For example, the influence of AL has been shown significant in randomally perturbed SLs~\cite{YanWang2014,luckyanova2018phonon,juntunen2019anderson,RenjiuHu2021} and NPCs~\cite{hu2018randomness}. However, this requires coherent phonon transport to persist over longer distances, and is still limited by the mobility edge frequency. Conversely, since resonance hybridizations influence arbitrary low frequencies, the size of the unit cell in an NPM does not set a lower limit for resonant localization in general. 
For the small-scale NPM models considered in this work (unit cell size of 3.26 nm), influencing the subwavelength regime enables the localization of low-frequency heat-carrying phonons, which for room-temperature silicon may include  wavelengths up to $\sim$10 nm~\cite{Chen2011}. For longer wavelength phonons that do not accommodate heat capacity, subwavelength resonant localization may take place with arbitrarily large NPM unit cells limited only by the phonon mean free path. Emergence of localized modes in the low-frequency range can indirectly affect key thermal properties relevant at the nanoscale, such as the non-local length and the heat flux relaxation time$-$which are influenced by a broader distribution of phonons compared to the range that directly influences the thermal conductivity~\cite{sendra2021}.~Another key aspect is that the emergence of phonon interference patterns in AL requires certain stochastic evolution of the non-equilibrium phonon population. In transient conditions at short time scales, such as in the wave-packet simulations, this is thus another limiting factor for energy localization via stochastic interaction with disorder. In contrast, local resonances are established deterministically and instantaneously in NPMs. From this perspective, the presence of local resonances offers a more effective and powerful mechanism than AL for phonon localization.



\begin{figure}
    \centering
    \includegraphics{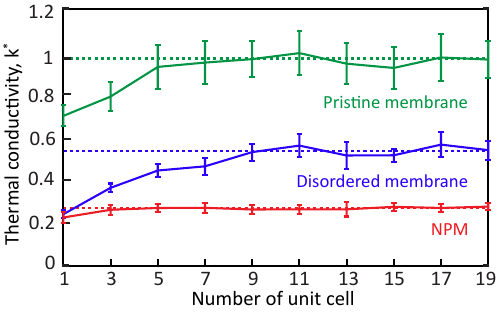}
    \caption{\textbf{Thermal conductivity} obtained from equilibrium MD simulations and the G-K method as a function of the number of unit cells considered in the pristine membrane (green), the membrane with mass-disorder (blue), and the NPM (red). All curves are normalized with respect to the converged themal conductivity value of the pristine membrane. The converged values are marked with dashed horizontal lines.}
    \label{fig:FigA1}
\end{figure}

\section{Conclusion}\label{conclusion}

We have demonstrated that the presence of atomic-scale local resonances within an anharmonic scattering environment offers an ordered counterpart to Anderson localization for phonons. While in AL coherent backscattering from disordered scattering centers impedes diffusion, in resonant localization diffusion is impeded by a deterministic resonant hybridization mechanism. Furthermore, while AL under a strong scattering environment is limited to relatively high frequencies and short wavelengths characterized by the mobility edge, resonant localization does not suffer from such a limit$-$the local resonances may be induced at exceedingly low frequencies impacting phonons with long wavelengths as long as the nanoresonator features are relatively small compared to the average phonon mean free path.~Within this mean-free path constraint, the strength of the localization may be increased by intrinsically increasing the number of local resonances, in analogy to increasing the degree of disorder in AL. \\
\indent Using frequency-dependent wave-packet excitations introduced into equilibrium MD simulations, where both harmonic and anharmonic effects are active, we have demonstrated explicitly this new phonon localization phenomenon. Parallel simulations of disordered lattices were also conducted to provide direct comparison between the two phonon localization mechanisms. Given that the observed resonant localization phenomenon takes place at discrete narrowband frequency ranges spanning the full spectrum, and highly tunable by design, the impact of the mechanism accumulates leading to fundamental changes to the overall character of the phonon transport. This has implications on the thermal conductivity and, more generally, the broader scope of phonon-related condensed matter phenomena 
such as phonon-electron couplings and quantum coherence~\cite{lilia2022,giustino2021}.

\section{Appendix}

\subsection{Methods}\label{methods}

\subsubsection{Nanostructures geometry and boundary conditions}\label{methods_geometries}

Each system geometry consists of a set of silicon conventional cells (CC) forming a larger unit cell to represent the nanostructure under consideration. Graphical illustration of the unit cell in each case is shown in Fig.~\ref{fig:Fig1}. Here we define the different unit cells precisely in terms of the number of conventional cells along each direction, where each CC consists of 8 atoms in a cubic region with a lattice constant of $a=5.431 \AA$. The pristine and disordered membrane comprise $6\times6\times6$ CC with periodic boundary conditions in the in-plane directions, and free surfaces at the top and bottom boundaries. In the case of the disordered membrane, the mass of each atom is randomly chosen among 50 possible masses according to a Gaussian distribution with a relative standard deviation of 10\% around the nominal atomic mass of silicon. The NPC unit cell is formed from a pristine membrane unit cell with a region sized $4\times4\times6$ CC removed from its center. Finally, the NPM unit cell is formed from a pristine membrane with a nanopillar sized $4\times4\times6$ CC contiguously added over the center of its top surface.

\subsubsection{Phonon dispersion relations: Harmonic lattice dynamics and anharmonic spectral energy density}\label{methods_dispersion}

The room-temperature phonon band structures in Fig.~\ref{fig:Fig1} are obtained using the Stillinger-Webber potential~\cite{SWpotential}. We show linear harmonic LD dispersion curves, represented by solid lines. These were generated using the General Utility Lattice Program (GULP)~\cite{gale2003general}. To account for the influence of anharmonic interactions, we also show the phonon band structures extracted directly from equilibrium MD simulations. This latter calculation is produced by implementing an SED formulation, essentially applying a Fourier transform to the MD velocity field to obtain a frequency-versus-wavenumber mapping of the energy density along the $\Gamma$X direction. We utilize an SED approach that requires knowledge of only the crystal unit-cell structure and does not require any prior knowledge of the phonon mode eigenvectors. The SED expression is a function of wave vector $\pmb{\kappa}$ and frequency $\omega$, and is given by~\cite{larkin2014comparison,thomas2010predicting}
\begin{equation}
{{{\Phi'}}\left( \pmb{\kappa}, \omega \right)} = \mu_{0} 
\displaystyle\sum\limits_{\alpha}^{3}
\displaystyle\sum\limits_{b}^{n}
\begin{vmatrix}
{\displaystyle\sum\limits_{l}^{N}{ \displaystyle\int\limits_{0}^{\tau_{0}}{\dot{{u}}_{\alpha}
\left (\scriptsize{\!\!\!\!
\begin{array}{l l} 
\begin{array}{l} l\\b
\end{array} \!\!\!\!\!\!
 &;~ t
\!\!\!\!\end{array}}
\right )
e^{{\textrm{i}}\left[\pmb{\kappa}\cdot\pmb{r}_{0}
\left (\!\!
\scriptsize{\begin{array}{l} l\\0
\end{array} \!\!}
\right )-wt\right]} \textrm{d} t}} }
\end{vmatrix}
^{2} ,
\label{eq:SED_Jonas}
\end{equation}
where ${\dot{{u}}}_{\alpha}$ is the $\alpha$-component of the velocity of the $b$th atom in the $l$th unit cell at time $t$, $\mu_{0}=m/(4\pi \tau_{0} N)$, $\tau_{0}$ is the total simulation time, and ${\pmb{r}}_{0}$ is the equilibrium position vector of the $l$th unit cell. There are a total of $N=N_{x}\times N_{y}\times N_{z}$ unit cells in the simulated computational domain with $n$ atoms per unit cell. We note that in Eq.~(\ref{eq:SED_Jonas}), the phonon frequencies can be obtained only for the set of allowed wave vectors as determined by the crystal structure. For our model, the $\Gamma X$-path wave vectors are $\kappa_{x}={2\pi j}/({N_{x}6a})$, $j=0$ to $N_{x}/2$. For the computational domain, we set $N_{x} = 50$ and $N_{y}= N_{z}=1$, which gives a $\Gamma X$ wave-vector resolution of ${\Delta}{\kappa}_{x}=0.02 (2\pi/ 6a)$. The MD simulations are performed using the LAMMPS software \cite{PLIMPTON19951}. Each system is simulated under $NVE$ ensemble at 300 K for $2^{22}$ time steps of size ${\Delta}{t}=0.5$ fs. Equation~(\ref{eq:SED_Jonas}) is evaluated on the velocity trajectories extracted every $2^{5}$ steps.

\begin{figure} [b]
    \includegraphics{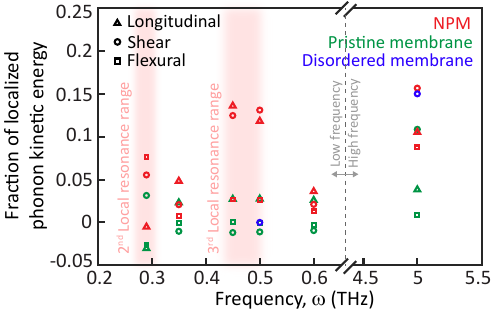}
    \caption{\textbf{Fraction of localized energy} as a function of excitation frequency within the central 8.2-nm wide slice averaged during the time window from 12 to 20 ps after excitation. The different symbols denote different force polarizations (triangles for longitudinal, circles for shear, and squares for flexural), and the different colors denote the different systems (green for pristine membrane, blue for membrane with mass-disorder, and red for NPM). The red regions denote the frequency ranges wherein resonant modes are identified in the NPM SED band structure.}
    \label{fig:Fig-fraction}
\end{figure}
\begin{figure*} 
    \centering
    \includegraphics{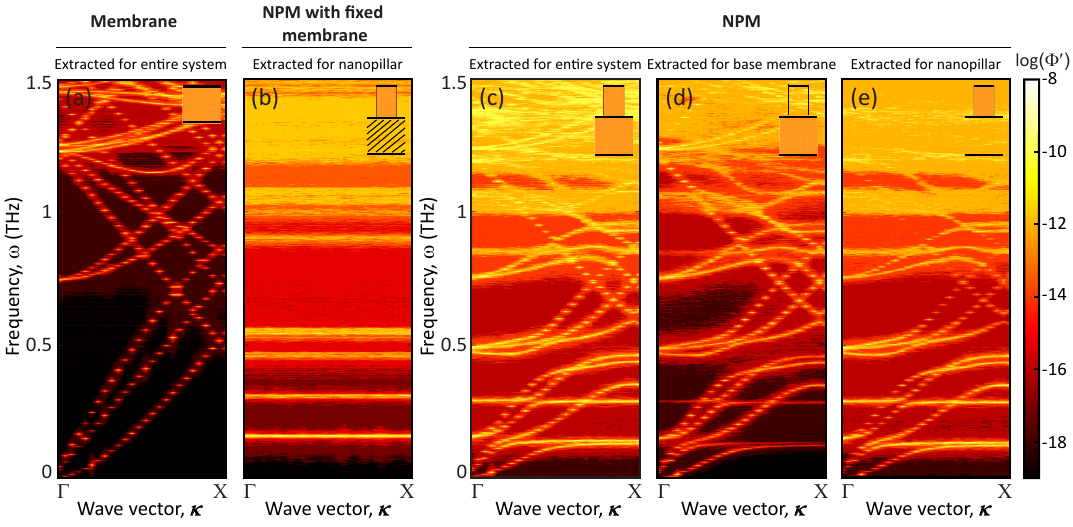}
    \caption{\textbf{Dissection of the membrane-nanopillar coupling in an NPM as illustrated in the SED phonon band structure.} (a) A uniform membrane with the same thickness of the NPM base membrane is also analyzed for comparison. (b) SED phonon band structure of the NPM with all base-membrane atoms fixed. SED phonon band of NPM considering (c) all atoms, (d) base membrane atoms only, and (e) nanopillar atoms only. As demonstrated in (d),  resonance hybridization is observed even when considering the motion of only the base membrane atoms in the NPM model. Schematics of unit-cell models shown in the insets. Solid black limes indicate a free surface.}
    \label{fig:FigA2}
\end{figure*}

\subsubsection{Wave-packet simulation method}\label{methods_WP}

To study wave-packet propagation in the different systems, we consider a super cell consisting of 50 aligned unit cells (see subsection \ref{methods_geometries} for a description of the unit cell in each case). This enables wave-packet propagation for long-enough distances to observe attenuation. The unit cells are aligned in the (1,0,0) crystalline orientation, which also corresponds to the direction of the wave-packet propagation. This choice prevents inhomogeneous energy injection within the excitation region. Similar to the band-structure calculations, the Stillinger-Webber potential is assumed in all the simulations \cite{SWpotential}.

\begin{figure}
    \centering
    \includegraphics{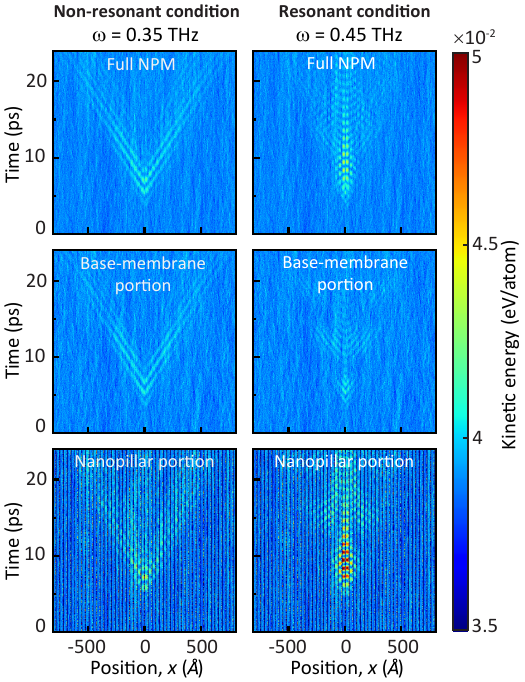}
    \caption{\textbf{Dissection of resonant localization phenomenon in NPM as illustrated in the wave-packet MD simulations.} Under a non-resonant condition, most of the wave-packet energy propagates through the base membrane in the NPM. For resonant frequencies, a larger fraction of the energy is stored in the nanopillars, however resonant phonon localization is also observed in the base-membrane portion.}
    \label{fig:FigA3}
\end{figure}

\begin{figure*} 
    \centering
    \includegraphics{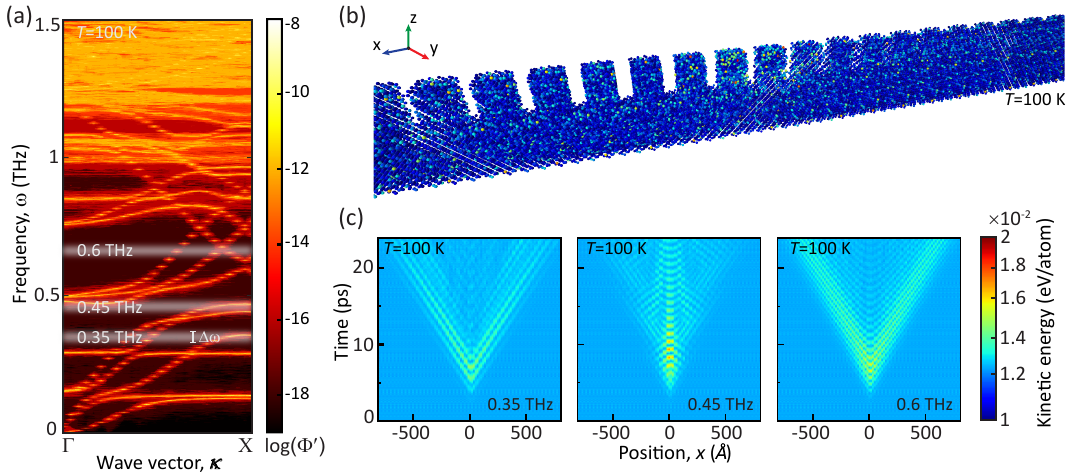}
    \caption{\textbf{Wave-packet propagation at 100 K.} (a) SED phonon band structure. (b) Kinetic energy in each atom at a representative time step after excitation  with $\omega=0.45$ THz. c) Kinetic energy density distribution in space and time after excitation at different frequencies.}
    \label{fig:FigA4}
\end{figure*}

The samples are first equilibrated for 100 ps at 300 K using a Nosé-Hoover thermostat. Then, the thermostat is switched off and the force excitation described by equation \eqref{eq:forceeq} is applied to each atom within a region forming a cuboid at the center of the domain that covers the full thickness along $z$ direction, the full width along the $y$ direction, and 4 Angstroms along the $x$ direction (see coordinate system in Figs.~\ref{fig:Fig2} and \ref{fig:Fig3}). In the NPM case, the excitation is performed in between nanopillars. The work performed by the excitation shown in Fig. \ref{fig:Fig5} is averaged over 10 independent simulations in each case. While the trajectory integral of the atomic-scale motion and the energy increase as described by Eqs.~\eqref{eq:work} and \eqref{eq:work2}, respectively, provide very similar average values for $W$, we obtained a smaller variance using the energy calculation. Moreover, the wave-propagation spatio-temporal heat maps shown in Figs. \ref{fig:Fig2}, \ref{fig:Fig3}, \ref{fig:Fig6}, \ref{fig:FigA3} and \ref{fig:FigA4} are obtained by averaging the kinetic energy of the atoms within 16 $\AA$-wide slices considering 10 simulations in each case. 

\subsubsection{Thermal conductivity prediction from equilibrium MD simulations and the Green-Kubo method}\label{methods_Green-Kubo}

We predict and compare the thermal conductivity in the different systems using equilibrium MD simulations followed by application of the G-K method~\cite{Schelling2002}. The systems are initially equilibrated for 1 ns using a time step of size $\Delta{t}=0.5$ fs under the $NPT$ ensemble at 300 K. Then, the heat flux integrated over the volume $\textbf{J}$ is sampled every 4 fs under the $NVE$ ensemble for 6 ns. The resulting flux autocorrelation function $<\textbf{J}(0)\otimes \textbf{J}(t)>$ is used to compute the thermal conductivity

\begin{equation}
     k=\frac{1}{k_\text{B}V T^2}\int <\textbf{J}(0)\otimes \textbf{J}(t)>\mathrm{d}t,
     \label{kGK_Jonas}
\end{equation}
where $T$ and $V$ are the average temperature and the volume of the entire material system, respectively. The results are averaged over 20 independent simulations.

The thermal conductivity $k$ is influenced by computational size effects \cite{Schelling2002}. Therefore, we compute $k$ as a function of the number of unit cells in each system (unit-cell sizes and geometries detailed in section \ref{methods_geometries}).~These results are shown in normalized form in Fig.~\ref{fig:FigA1} where the converged values for $k$ are marked by horizontal dashed lines. We note that a larger number of unit cells is required to converge the thermal conductivity values in the disordered membrane than in the pristine systems. While in the latter cases all the unit cells are equivalent, proper sampling of the randomly-varied mass of the atoms in the disordered membrane requires a significantly larger system size.

\subsection{Fraction of Localized Energy}\label{localized_energy_fraction}

Here we provide a broader comparison of the degree of wave-packet energy localization obtained in the different systems by varying the frequency and the polarization of the excitation. In Fig.~\ref{fig:Fig-fraction}, we plot the fraction of non-equilibrium kinetic energy that remains within a 8.2 nm-width slice centered in the excited region and averaged over the time-window from 12 ps to 20 ps after excitation.~This magnitude is obtained by subtracting the background energy before excitation from the total energy measured during the wave-packet simulation. We show normalized results with respect to the total energy injected by the external force, and averaged over 10 independent realizations.~The resulting fraction values in Fig.~\ref{fig:Fig-fraction} are strongly dependent on the arbitrary choices for the spatio-temporal measurement window.~In addition, the spontaneous energy fluctuations around the background energy induce some noise in the characterization of this metric, thus resulting in negative values for some cases where there is no localization. However, the results provide equal-footing comparison of the localization effect across the different systems, polarizations, and frequencies.\\
\indent At low frequencies, all the input energy departs from the excitation region during the initial 12 ps in the pristine and disordered membranes for any polarization, thus the fraction of localized energy is zero. In contrast, a significant fraction of energy remains localized in the NPM system for frequency values around the resonances. In particular, we observe a larger fraction of localized energy in the NPM system relative to the membrane at $\omega=0.29$ THz, which coincides with a resonant mode in the subwavelength regime (i.e., with wavelengths larger than the NPM unit cell size along the direction of transport). Since the exact resonant frequencies and the kinetic energy stored by the corresponding modes are not equivalent for different polarizations, we observe different results for forces polarized differently. Finally, at high frequencies, a significant amount of energy remains in the excited region even for the pristine membrane due to the small velocity of the corresponding phonon modes. However, localization effects further limit the propagation of kinetic energy in the presence of disorder or resonances.\\
\indent As mentioned earlier, an attractive aspect of resonant localization is the ability to access exceedingly low frequencies, e.g., below the mobility edge that limits AL. This is beneficial for multiple reasons. First, it allows for a stronger reduction in the thermal conductivity (without the need to increase the unit-cell size) because low-frequency phonons contribute strongly to the thermal conductivity. Second, other average quantities beyond the thermal conductivity, such as the non-local length characterizing the heat flux correlations~\cite{sendra2021}, are determined by a broader spectrum than the thermal conductivity. Third, other phenomenon, such as phonon-electron coupling for example, are also generally influenced by a broader portion of the spectrum including low frequencies.\\
\indent Another favorable feature of resonant localization is its high degree of tunability. It is possible to determine \textit{a priori} the precise spatial distribution of the localization by controlling the locations where the nanoresonators are placed within the crystal. In addition, the target frequencies for localization exactly correspond to the resonances of the nanopillar, which can be explicitly computed. 

\subsection{Membrane-nanopillar couplings in NPM}\label{memb-pillar_coupling}

In Section~\ref{deterministic} and~\ref{work_min}, we demonstrated localization of wave-packet induced phonon kinetic energy and minimization of the external work at the specific frequencies where mode hybridizations are identified in the anharmonic dispersion relation of the NPM. This is a consequence of the coherent interaction between the vibrational modes in the nanopillars and the propagating phonon modes in the underlying base membrane. 

In Fig.~\ref{fig:FigA2}, we investigate the phonon-vibron coupling phenomena further by analyzing the NPM anharmonic phonon band structure when considering exclusively the atomic motion of different spatial subregions within the NPM unit cell. As two reference cases, we also show the SED dispersion for a standard uniform membrane [Fig.~\ref{fig:FigA2}(a)], where no mode hybridizations occur, and for an NPM model where all the atoms in the basis membrane are fixed (for demonstrative purposes) [Fig.~\ref{fig:FigA2}(b)]. In the latter case, we observe clearly isolated and distinct $\pmb{\kappa}$-independent curves (in the form of horizonal lines), each representing the nanopillar vibrations (or standing waves) at the various resonance frequencies. These resonance curves are numerous (totalling the number of atoms in the nanopillar multiplied by three), are distributed across the full spectrum, and depend only on the nanopillar geometric properties. These two results together represent a ``dissection" of the modal features of the standard NPM model (with no atoms fixed). The standard NPM is shown in Fig.~\ref{fig:FigA2}(c) where the dispersion displays effectively a merger of the constituent band structures of Figs.~\ref{fig:FigA2}(a) and~\ref{fig:FigA2}(b). Specifically, the core formation of both the membrane phonon dispersion curves and the resonance flat curves are intact, but the two sets couple and hybridize which leads to a foundational change in the phonon, and consequently thermal transport, properties of the overall system. Each of these couplings form an avoided crossing where the strength of the repulsion between the curves increases with the strength of the coupling. Furthermore, we note that the shapes of the dispersion curves in the full NPM system are almost identical to the ones considering only the dynamics of the atoms in the base membrane [Fig.~\ref{fig:FigA2}(d)], or the atoms in the nanopillar [Fig.~\ref{fig:FigA2}(e)], respectively. This illustrates the strong coupling between the degrees of freedom of the nanopillar and the base membrane, suggesting that the modal character is encoded in \text{all} atoms of the NPM unit cell. This is the case when the NPM feature sizes are smaller than the average MFP. This condition is indeed the key to the realization of the deterministic phonon localization mechanism identified in pristine NPMs.\\
\indent The implications of the membrane-nanopillar coupling is further investigated by analyzing the downstream spatial evolution of the wave-packet kinetic energy distribution within different regions of the NPM structure. In Fig.~\ref{fig:FigA3}, we show the spatio-temporal evolution of the kinetic energy within the full NPM domain (i.e., comprising both the base membrane and the attached nanopillar) and within only the nanopillar or the base membrane regions, respectively.~We again consider both non-resonant and resonant excitation conditions. In the former case, the kinetic energy propagation within the base membrane is very similar to that in the full NPM structure. This indicates that the coherent coupling between nanopillar and membrane is not dominant at this frequency. Conversely, under resonant conditions, a smaller fraction of the kinetic energy propagates through the base membrane because a large amount of wave-packet kinetic energy gets stored in the nanopillars. What is important, however, is that the remaining energy that is accommodated in the base membrane is clearly localized within the first few NPM unit cells along the direction of transport. This further demonstrates that the presence of the nanopillars is influencing the transport in the base membrane portion beyond a single unit cell. \\
\indent The SED and wave-packet simulation results shown in Fig.~\ref{fig:FigA2} and \ref{fig:FigA3} account for anharmonic effects in room-temperature silicon through the empirical Stillinger-Webber interatomic potential. Therefore, we conclude that resistive phonon-phonon scattering does not significantly attenuate the coherent interaction of the phonons and the resulting non-local effects pertaining to its wave-like properties in our simulations. \\

\subsection{Wave-packet propagation at low temperatures} \label{low_temperatures}

All the results above are obtained at 300 K to emphasize that the phenomenon of resonant localization of phonons at the nanoscale is feasible for room-temperature applications, and potentially at higher temperatures as well. In Fig.~\ref{fig:FigA4}, we show that similar wave-packet kinetic energy localization at the resonant frequency is obtained by reducing the temperature to 100 K. In this case, the coherent response is expected to persist more profoundly and for longer time windows due to the diminishing of anharmonic effects at lower temperatures (i.e., due to the the increase in the phonon mean free values for all modes). Moreover, as also displayed in Fig.~\ref{fig:FigA4}, the SED anharmonic phonon structure is only slightly modified relative to the room-temperature case, and hence the resonant phonon modes are identified almost at the same frequencies as in higher temperatures.

\bibliographystyle{naturemag}
\bibliography{biblio}

\end{document}